# Segmentation of the prostate and organs at risk in male pelvic CT images using deep learning

Samaneh Kazemifar*, Anjali Balagopal*, Dan Nguyen, Sarah McGuire, Raquibul Hannan, Steve Jiang, Amir Owrangi

Medical Artificial Intelligence and Automation Laboratory, Department of Radiation Oncology, University of Texas Southwestern, Dallas, Texas, US

Emails: Amir.Owrangi@UTSouthwestern.edu, Steve.Jiang@UTSouthwestern.edu

**Running Title:** Organ segmentation in male pelvic CT images

* Co-first authors

**Abstract:**

Inter-and intra-observer variation in delineating regions of interest (ROIs) occurs because of differences in expertise level and preferences of the radiation oncologists. We evaluated the accuracy of a segmentation model using the U-Net structure to delineate the prostate, bladder, and rectum in male pelvic CT images. The dataset used for training and testing the model consisted of raw CT scan images of 85 prostate cancer patients. We designed a 2D U-Net model to directly learn a mapping function that converts a 2D CT grayscale image to its corresponding 2D OAR segmented image. Our network contains blocks of convolution 2D layers with variable kernel sizes, channel number, and activation functions. On the left side of the U-Net model, we used three 3x3 convolutions, each followed by a rectified linear unit (ReLu) (activation function), and one max pooling operation. On the right side of the U-Net model, we used a 2x2 transposed convolution and two 3x3 convolution networks followed by a ReLu activation function. The automatic segmentation using the U-Net generated an average dice similarity coefficient (DC) and standard deviation (SD) of the following: DC ± SD (0.88 ± 0.12), (0.95 ± 0.04), and (0.92 ± 0.06) for the prostate, bladder, and rectum, respectively. Furthermore, the mean of average surface Hausdorff distance (ASHD) and SD were 1.2 ± 0.9 mm, 1.08 ± 0.8 mm, and 0.8 ± 0.6 mm for the prostate, bladder, and rectum, respectively. Our proposed method, which employs the U-Net structure, is highly accurate and reproducible for automated ROI segmentation. This provides a foundation to improve automatic delineation of the boundaries between the target and surrounding normal soft tissues on a standard radiation therapy planning CT scan.

Keywords: Organ segmentation, Prostate, CT images, Deep learning, U-Net architecture

# 1. Introduction

Current variability in manual structure delineation of computed tomography (CT) images in radiation therapy introduces more errors than daily setup uncertainties[1-6]. Large inter-and intra-observer variation in delineating the regions of interest (ROIs) has been reported because of differences in level of expertise and preferences of the radiation oncologists[7-10]. Steenbergen *et al.*[11] reported substantial inter-observer variation in manual prostate delineation. Delineation variation between different observers reflects the uncertainty in defining this boundary, and is especially pronounced near the tumor boundary. Manual delineation of the ROIs by visual inspection is considered the gold standard in current clinical practice despite being time consuming and not especially robust, indicating the need for a more accurate, reliable, and robust segmentation method. With the use of intensity modulated radiation therapy (IMRT), where steep gradients sculpt the dose away from the organs at risk (OARs), accurate delineation becomes essential to avoid situations of large geometric miss that lead to higher dose to the target. Also, accurate delineation is critical to prevent recurrences. Highly precise dose delivery to a poorly delineated target negates any benefits of technology improvements because of low accuracy[12]. No level of on-board image guidance will eliminate these systematic delineation errors[12]. The consequences of systematic uncertainties can be severe, including acute or late Grade 2 rectal toxicity in more than 6% of patients and acute or late Grade 2 urinary toxicities in 37% of patients using IMRT plans developed from contours drawn on CT images[13]. However, toxicities can be reduced substantially if the ROI is delineated more accurately either through imaging techniques with higher soft tissue contrast such as MRI[14] or computer-aided techniques from CT images.

Automatic segmentation algorithms can help radiation oncologists delineate ROIs more accurately, consistently, and efficiently[15-18]. Classic computerized image segmentation methods are based on image intensity, gradient, and texture; however, delineating accurate boundaries between the OARs and target based on gray-scale intensity value differences is either challenging or unfeasible because of the low contrast to noise ratio (CNR) of low soft tissue in CT images. Although secondary imaging modalities such as MRI and PET are often used to distinguish between tumor and OARs, the resulting images are often unavailable, and also must be registered to a CT for accurate dose calculation. This introduces registration errors from set-up variations during each imaging session and presents inherent limitations in accurately matching images with different information[19,20]. Therefore, generating accurate OARs and tumor boundaries from a CT image is highly desirable to minimize treatment uncertainties in radiation oncology.

The automated medical image segmentation methods proposed so far can be categorized into intensity-based[21] and texture-based[22]. Both approaches are sensitive to noise when selecting optimal threshold values. Also, the probability of over or under segmentation affects the performance of these methods. Other methods include model-based and atlas-based techniques. Model-based methods[23-27], such as level-set techniques, are based on shape and appearance of the object and require an initial guess of the parameters. This may be suboptimal especially in tumor/OAR boundary identification because tumor shape is unpredictable and may affect the shape of neighboring OARs. On the other hand, atlas-based approaches[28-30] are available in some clinical treatment planning systems, but face many limitations including the selection of optimal atlases[31,32] and a true representation of the study population, which may also be affected by the presence of tumor tissue. Some semi-automated learning-based methods[33-36] for prostate segmentation have performed remarkably well in recent years. In such methods, each voxel is labeled explicitly according to the target object or background in the CT image. The semi-automated prostate segmentation method proposed by Shi *et al.*[36] classifies the pixels by applying a spatially-constrained transductive lasso on local region features to select joint features. Gao *et al.*[33] proposed a displacement regressor that predicts 3D displacement to assist learning of the above classifier for accurate pelvic organ segmentation. Shao *et al.*[35] presented a boundary detector based on a regression forest and used it as an initial shape before guiding accurate prostate segmentation. Lay *et al.*[34] proposed a discriminative classifier by employing the landmarks that can be detected through joint global and local texture information. The drawback of these methods is that they depend on predefined features, affecting the accuracy of segmentation.

Despite progresses achieved in the field of organ segmentation, there is a critical need to bridge the gap between automated segmentation results and manual annotations. The challenges are largely related to variability in size, shape, and contour of the ROI, and can be resolved by applying methods that use *a priori* knowledge such as machine learning techniques. Machine learning methods such as deep learning models have been developed in the field of computer vision[37] and have become the state-of-the-art in many applications[38,39]. Feature-driven model-based methods[40-42] using machine learning techniques have been proposed. Also, appropriate features are extracted and patterns are learned through conventional machine learning methods. Deep learning methods discover the features in a hierarchical fashion instead of using handcrafted features based on an initial guess. In other words, deep learning methods learn the low-level features first, and more comprehensive high-level features on a layer by layer basis later. Two recently published studies showed the segmentation of OARs on CT images using deep learning methods. Ibragimov and Xing[43] described segmenting OARs in CT images of the head and neck using standard convolutional neural networks (CNNs). Men *et al.*[44] illustrated the benefits of using a deep dilated CNN (DDCNN) to segment the clinical target volume (CTV) and OARs for rectal cancer using CT data. The goal of image segmentation in biomedical image processing is to label each pixel of an image to a certain class. In the deep learning scope, the fully convolutional network proposed by Long *et al*[45] is a benchmark of image segmentation. Pixel level segmentation is conducted by replacing the fully connected layer with the convolution layer. However, this replacement generates a coarse segmentation map because information is lost during the pooling operation. Three other structures were developed to address this problem, including dilated convolution[46], encoder-decoder convolutional network[47], and U-Net model[48], which may improve the resolution of the segmentation results. In the encoder-decoder technique, the advantage of the structure is to connect the pooling layers with the unpooling layers. In U-Net, the usual contracting network in CNN is supplemented by successive layers, where pooling operators are replaced by up-sampling operators. These layers increase the resolution of the output, and high resolution feature maps from the contracting path are combined with the corresponding up-sampling feature maps in the expansive path, enabling us to fully use the limited data and largely address the overfitting problem. The goal of this study is to evaluate the accuracy and robustness of a segmentation model using the U-Net structure to delineate the prostate, bladder, and rectum in male pelvic CT images.

## 2. Methods

## 2.1 Study Participants

The dataset used for training and testing the model consisted of raw CT scan images of 85 prostate cancer patients collected at the University of Texas Southwestern Medical Center (UTSW). All CT images were acquired using a 16-slice CT scanner (Royal Philips Electronics, Eindhoven, The Netherlands). The target organ was the prostate and OARs included the bladder and rectum. All contours were drawn by radiation oncologists. All images were acquired with a 512x512 matrix and 2-mm slice thickness (voxel size 1.17mm×1.17mm×2mm). CT images and their corresponding structural images were randomly assigned to either training (70%) or testing (30%) sets.

## 2.2 Deep CNN Model

We designed a 2D U-Net[48] model to directly learn a mapping function that converts a 2D CT grayscale image to its corresponding 2D OAR segmented image. The model can be learned using 2D CT images with corresponding 2D OAR segmentation from each training participant as the input. Once the model is trained, it can be applied on an unseen CT image (test data) to segment the final OARs and prostate image slice by slice. The outline of the proposed U-Net network for male pelvic image segmentation is shown in Figure 1. Generally, the U-Net model includes convolution (contracting path) and deconvolution (expansive path) networks for pixel-wise predictions and in this network, input and output image sizes are the same. The left side of the U-Net model contains the repetition of the convolutional network, and the right side contains an up-sampling of the feature map followed by the convolutional network and a concatenation with the feature map from the left side. Our network contains blocks of convolution 2D layers with variable kernel sizes, channel numbers, and activation functions. On the left side, we used three 3x3 convolutions, each followed by a rectified linear unit (ReLu) (activation function), and one max pooling operation. On the right side, we used a 2x2 transposed convolution and two 3x3 convolution networks followed by a ReLu activation function. In addition, batch normalization[49] and dropout[50] were added to the layers. In the final layer, we used a 1x1 convolution network with sigmoid activation function and dice similarity coefficient (DC) loss function. The stochastic gradient descent (SGD) optimizer used a learning rate of 0.01 and a momentum of 0.9 to update network weights iteratively based on training data. Details of the U-Net structure are shown in Table 1. The U-Net model was implemented using the open source Keras package[51], and componential processing was performed with a NVIDIA Tesla K80 dual-GPU graphic card. A preprocessing step was added to crop the CT/ROIs pair images into 128x128 images for the rectum and prostate, and 160x160 for the bladder from the original 512x512 images, while maintaining the original image resolution. This step is needed to overcome the memory limitation during model training using the NVIDIA Tesla GPU card.

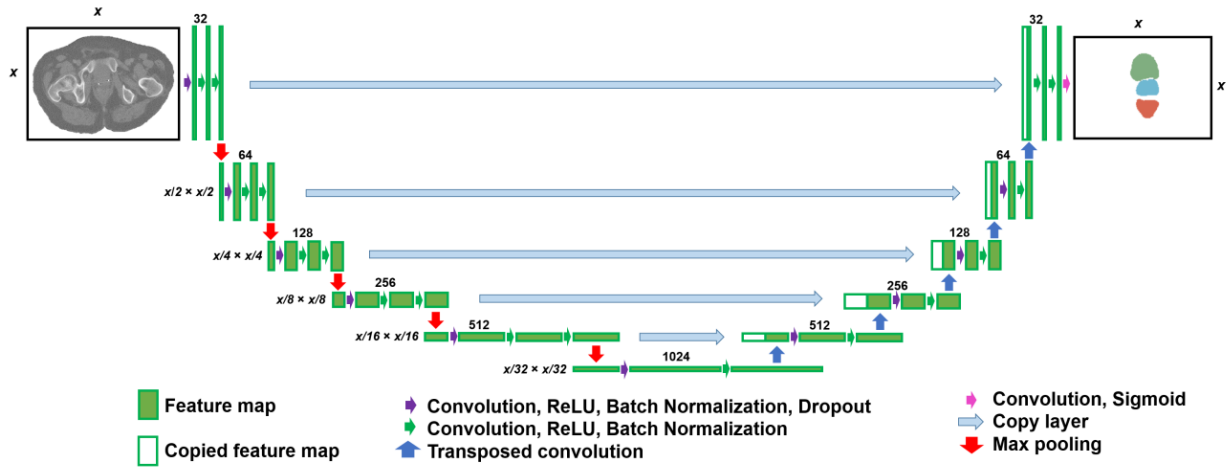

**Figure 1.** Block diagram of the U-Net structure for male pelvic CT images segmentation. The left side of the unit shows the convolution part and the right side shows the deconvolution part. Green arrows denote 2D convolution layer with kernel size (3,3). Red arrows indicate max pooling layers with size (2,2). Blue arrows show the con2Dtranspose with kernel size (2,2). Batch-normalization and dropout layers are applied after each conv2D layers. ROIs include prostate, rectum, and bladder. X indicates the size of the images.

**Table 1** Layers and parameters of the U-Net model for male pelvic segmentation in CT images.

| Layers | Parameters | Activation |
|---|---|---|
| Dropout, BatchNormalization, Conv2D | 3x3x32; dropout rate=0.2; padding | ReLu |
| BatchNormalization, Conv2D | 3x3x32; padding | ReLu |
| BatchNormalization, Conv2D | 1x1,32,padding | ReLu |
| Maxpooling1 | 2x2 | |
| Dropout, BatchNormalization, Conv2D | 3x3x64; dropout rate=0.2;padding | ReLu |
| BatchNormalization, Conv2D | 3x3x64; padding | ReLu |
| BatchNormalization, Conv2D | 1x1x64; padding | ReLu |
| Maxpooling2 | 2x2 | |
| Dropout, BatchNormalization, Conv2D | 3x3x128; dropout rate=0.3; padding | ReLu |
| BatchNormalization, Conv2D | 3x3x128; padding | ReLu |
| BatchNormalization, Conv2D | 1x1x128; padding | ReLu |
| Maxpooling3 | 2x2 | |
| Dropout, BatchNormalization, Conv2D | 3x3x256; dropout rate=0.3; padding | ReLu |
| BatchNormalization, Conv2D | 3x3x256; padding | ReLu |
| BatchNormalization, Conv2D | 1x1x256; padding | ReLu |
| Maxpooling4 | 2x2 | |
| Dropout, BatchNormalization, Conv2D | 3x3x512; dropout rate=0.4; padding | ReLu |
| BatchNormalization, Conv2D | 3x3x512; padding | ReLu |
| BatchNormalization, Conv2D | 1x1x512; padding | ReLu |
| Maxpooling5 | 2x2 | |
| Dropout, BatchNormalization, Conv2D | 3x3x1024; dropout rate=0.4; padding | ReLu |
| BatchNormalization, Conv2D | 3x3x1024; padding | |
| Concatenate, Con2DTranspose | 2x2x512; stride:2x2; padding | |
| Dropout, BatchNormalization, Conv2D | 3x3x512; dropout rate=0.4; padding | ReLu |
| BatchNormalization, Conv2D | 3x3x512; padding | ReLu |
| Concatenate, Con2DTranspose | 2x2x256; stride:2x2; padding | |
| Dropout, BatchNormalization, Conv2D | 3x3x256; dropout rate=0.4; padding | ReLu |
| BatchNormalization, Conv2D | 3x3x256; padding | ReLu |
| Concatenate, Con2DTranspose | 2x2x128; stride:2x2; padding | |
| Dropout, BatchNormalization, Conv2D | 3x3x128; dropout rate=0.4; padding | ReLu |
| BatchNormalization, Conv2D | 3x3x128; padding | ReLu |
| Concatenate, Con2DTranspose | 2x2x64; stride:2x2; padding | |
| Dropout, BatchNormalization, Conv2D | 3x3x64; dropout rate=0.4; padding | ReLu |
| BatchNormalization, Conv2D | 3x3x64; padding | ReLu |
| Concatenate, Con2DTranspose | 2x2x32; stride:2x2; padding | |
| Dropout, BatchNormalization, Conv2D | 3x3x32; dropout rate=0.4; padding | ReLu |
| BatchNormalization, Conv2D | 3x3x32; padding | ReLu |
| Conv2D | 1x1x1 | sigmoid |

## 3. Results

The segmentation results for prostate, bladder, and rectum using the U-Net model are illustrated in Figure 2 and Figure 3. For better visualization, we show only the part of the CT image with three ROIs. As evident in Figure 2 and Figure 3, the prostate and the rectum are more difficult to segment than the bladder because of their unclear boundaries. The ground truth segmentation and the overlay map between both segmentations of the same subjects are also shown in Figure 2 and Figure 3. Our method using U-Net model can properly delineate the ROI boundaries when automatic and ground truth segmentation results overlap. However, the large shape variation of pelvic ROIs and the unclear boundaries between prostate and rectum is known in CT images. We used the average DC value and standard deviation between the segmentation results using our proposed method and ground truth segmentations to evaluate the accuracy of the U-Net model for the prostate, bladder, and rectum. Additionally, we used the average Hausdorff distance (AHD) (mm), the average surface Hausdorff distance (ASHD) (mm) and positive predictive value (PPV) to evaluate the segmentation performance as defined below:

$$AHD = d_{H,avg}(X,Y) = \frac{1}{|X|}\sum_{x \in X} \min_{y \in Y} d(x,y) \qquad (1)$$

$$ASHD = d_{H,avg}(X,Y) = \frac{d_{H,avg}(X,Y) + d_{H,avg}(Y,X)}{2} \qquad (2)$$

$$PPV = \frac{TP}{TP + FP} \qquad (3)$$

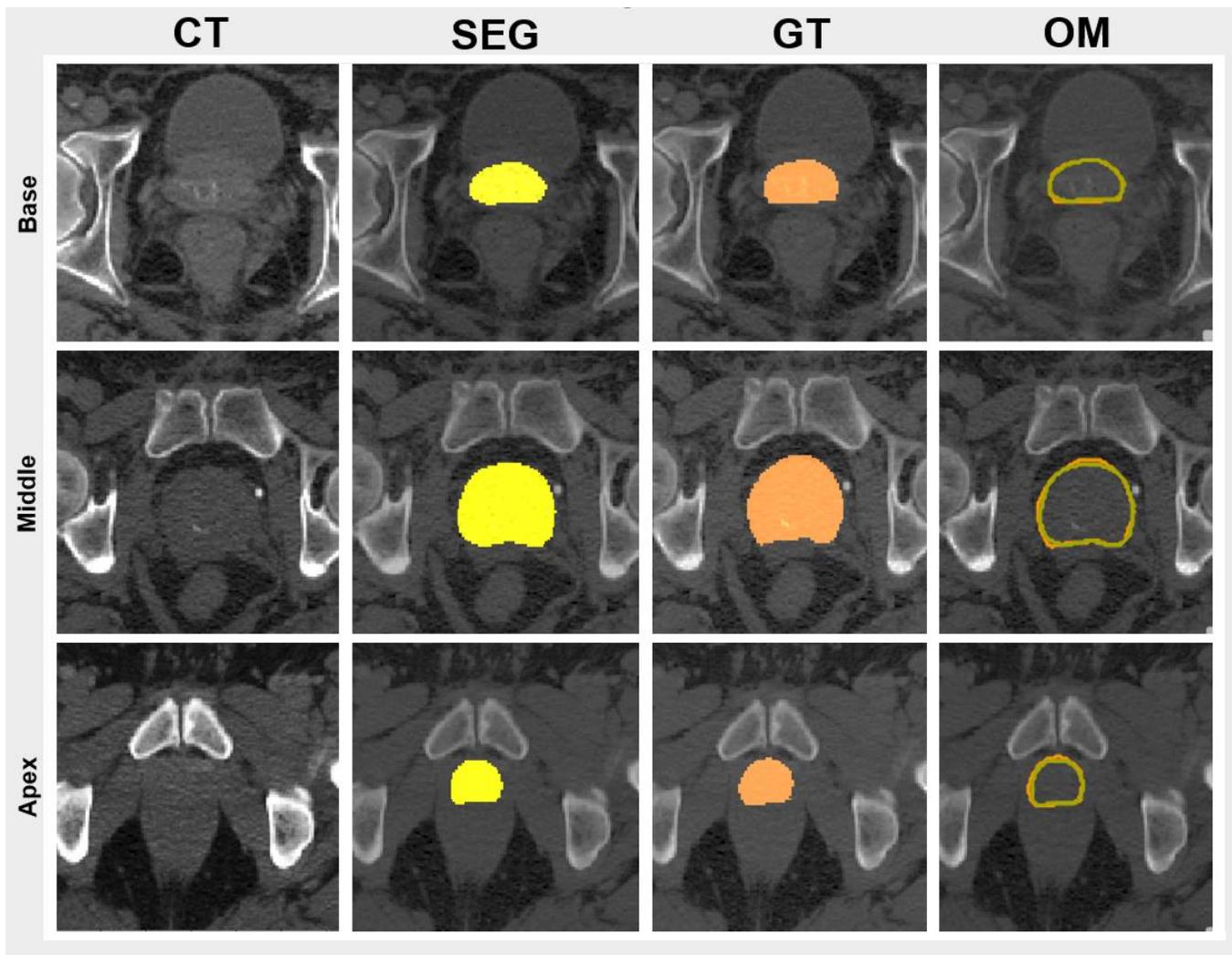

**Figure 2** Example of U-Net segmentation vs. ground truth results of a transverse CT image of the prostate at base, middle, and apex slices from top to bottom. From left to right, CT image (CT), U-Net segmentation (SEG), ground truth (GT), and overlay map (OM) of ground truth and U-Net segmentation images.

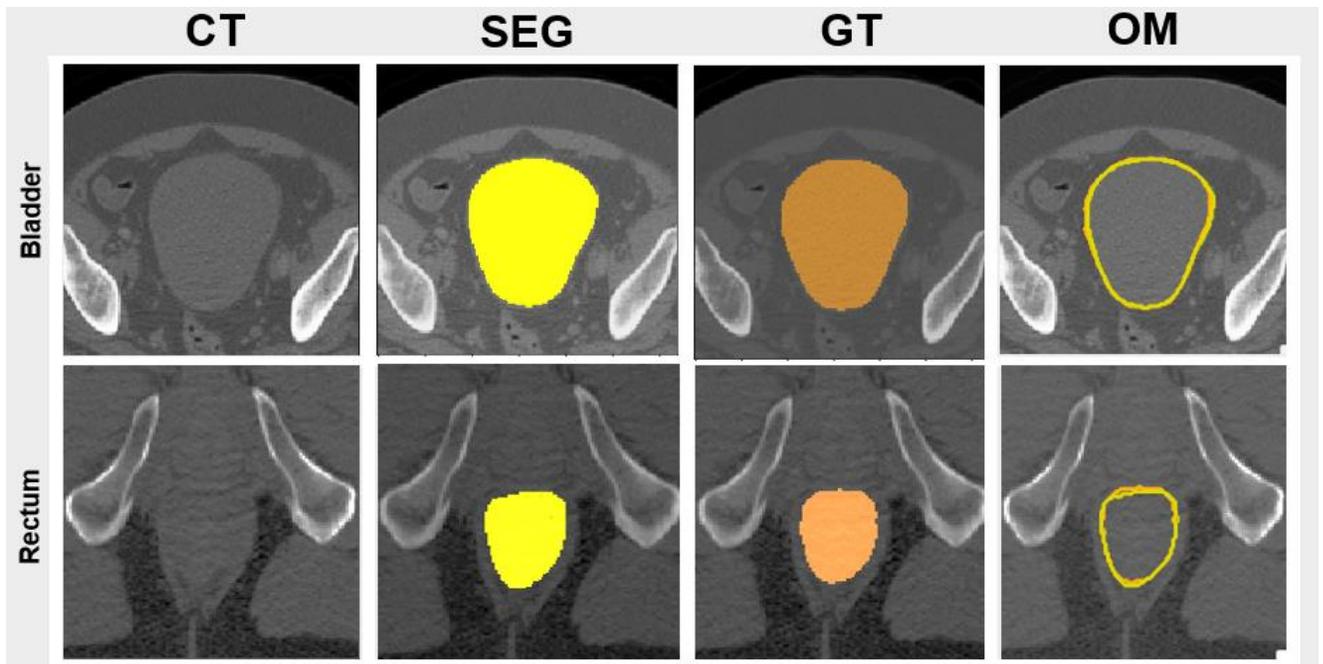

**Figure 3** Example of U-Net segmentation vs. ground truth results of the transverse CT image of the bladder and rectum from top to bottom. From left to right, CT image (CT), U-Net segmentation (SEG), ground truth (GT), and overlay map (OM) of ground truth and U-Net segmentation images.

where $X$ denotes the voxel set of ground truth volume, $Y$ denotes the voxel set of segmentation results, and $d(X, Y)$ is the Euclidean distance between $X$ and $Y$. TP, FP indicate the true positive rate and the false positive rate. The average and standard deviation values for each OAR using the proposed method are illustrated in Figure 4 and Table 2. Automatic segmentation using the U-Net model generated an average DC ± SD (0.88 ± 0.12), (0.95 ± 0.04), and (0.92 ± 0.06) for prostate, bladder and rectum, respectively. The AHD and ASHD for all three structures were on the order of 1 – 2mm. Considering the image voxel size is 1.17mm×1.17mm×2mm, these values are within the expected random error due to the image resolution limitations. Table 3 shows high overall sensitivity of the U-Net model. We obtained the highest average DC value, sensitivity and PPV for the bladder because of its high contrast, regular shape, and larger size compared to the rectum and prostate. As expected, the prostate had the lowest average DC value, sensitivity and PPV because this is the most irregular and uncertain shape to identify. These results indicate the effectiveness of using the U-Net model for segmenting critical tumor and normal tissue pelvic ROIs for radiation therapy.

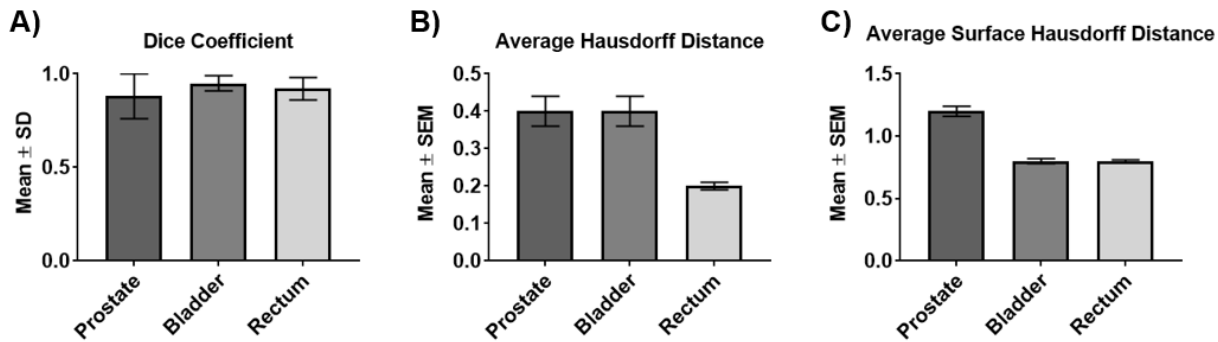

**Figure 4** A) Average and standard deviation (SD) of the dice similarity coefficient for prostate, bladder, and rectum. B) Average and standard error of mean (SEM) of the average Hausdorff distance (mm) for prostate, bladder, and rectum. C) Average and standard error of mean (SEM) of the average surface Hausdorff distance (mm) for prostate, bladder, and rectum using the U-Net model.

**Table 2** Mean and the standard deviation (SD) of different anatomic regions. Dice similarity coefficient (DC), average Hausdorff distance (AHD) (mm), and average surface Hausdorff distance (AHD) (mm) between U-Net segmentation and ground truth results for prostate, bladder, and rectum.

| ROI | DC ± SD | AHD ± SD | ASHD ± SD |
|---|---|---|---|
| **Prostate** | 0.88 ± 0.12 | 0.4 ± 0.7 | 1.2 ± 0.9 |
| **Bladder** | 0.95 ± 0.04 | 0.4 ± 0.6 | 1.1 ± 0.8 |
| **Rectum** | 0.92 ± 0.06 | 0.2 ± 0.3 | 0.8 ± 0.6 |

**Table 3** Mean of different anatomic regions. Sensitivity and positive predictive value (PPV) between U-Net segmentation and ground truth results for prostate, bladder, and rectum.

| ROI | Sensitivity | PPV |
|---|---|---|
| **Prostate** | 0.87 | 0.92 |
| **Bladder** | 0.95 | 0.95 |
| **Rectum** | 0.92 | 0.92 |

## 4. Discussion

Target and OARs segmentation is a critical step before optimizing radiation therapy treatment planning, but CT images do not provide strong soft-tissue contrast to delineate these structures. Accurate and reproducible automated ROI segmentation improves automatic delineation of the boundaries between the target and surrounding normal soft tissues on a standard radiation therapy planning CT scan. This could reduce normal tissue radiation toxicity while accurately controlling disease for patients with prostate cancer. Additionally, dose-volume relationships for prostate toxicity in patients could be understood better with reduced OAR delineation variation. A better understanding of the relationship between bladder and rectal toxicity and radiation dose would provide an opportunity to address tumor dose escalation and improve outcomes in patients with prostate cancer.

Our proposed application provides accurate and reproducible automated ROI segmentation using a deep CNN model, and increases consistency and robustness in delineating OARs. Deep CNN model training includes computation-intensive processes, but once the network is trained, the automated segmentations are computed for all ROIs within minutes. Automatic segmentation is faster than manual segmentation, which may take 20 - 30 minutes. This method could be easily applied to routine clinical practice of OAR and prostate segmentation in CT images to improve quality and consistency between patients and increase efficiency in the clinic. While experts take about 20 minutes to segment structures with poor CNR, the computational time in our study was about 1 second per patient to segment all three OARs using a NVIDIA Tesla K80 dual-GPU graphic card. The time taken to segment all the structures with the DDCNN structure developed by Men *et al* was ~ 45 second per subject using a Titan Z graphics card[44]. Training and testing time could be decreased by upgrading the GPU cards. The U-Net structure does not have input image size limitations, which makes the U-Net model a strong candidate for segmenting biomedical images. However, in this study we used cropped images because of memory problems, but in future studies we plan on improving the GPU cards to overcome memory limitation. Furthermore, we will work on improving the architecture of the U-Net model using 3D implementation[52] and a ResNet[53] block at the center of the U-Net structure. The advantage of the ResNet model is to ensure that the input properties of the early layers are available while developing the later layers, so that the output deviation can be controlled to decrease training time.

We used five down-sampling and five up-sampling blocks. In each block, the dropout layer was added to prevent overfitting during the training phase, gradually increasing the dropout rate in the down-sampling layers and gradually decreasing it in the up-sampling layers. Drop-out rates of 0.2 to 0.4 were used in the down-sampling section, and rates of 0.4 to 0.2 were used in the up-sampling section of the developed U-Net model. The average dice similarity coefficients were 0.95, 0.88, and 0.92 for the bladder, prostate, and rectum, respectively. These results indicate high similarity between automated and manual segmentations of the ROIs. Manual segmentation is challenging for these OARs because of inhomogeneity of the bladder and rectum, high intra- and inter-appearance variability, and poor boundary contrast between nearby organs. Other conventional automated segmentation methods such as single atlas-based techniques may present high registration errors because of large variability between the single atlas and target images. To overcome this problem, multi-atlas approaches were developed, but the selection of several templates based on similarity measurements is still challenging. Acosta *et al.*[54] compared bladder, rectum, and prostate CT image segmentation using different atlas rankings, selection strategies, and label fusion approaches such as the STAPLE method[55,56] or the vote algorithm. They showed that the average dice similarity coefficient values increased for the bladder (best average: ~0.92), rectum (best average: ~0.80), and prostate (best average: ~0.85) as a function of the number of atlases (between 1 and 29) used for two different label fusion rules. The dice similarity coefficient was increased for these regions by increasing the number of atlases, and the best results were achieved after including 20 atlases. Recently, Men *et al* applied a dilated CNN model to auto-segment the CTV and OARs for rectal cancer from CT images. They achieved 0.87, 0.93, 0.92, 0.92, 0.65, and 0.61 average dice similarity coefficients for the CTV, bladder, left and right femoral heads, rectum and bowel, respectively. We reported several statistical measurements (Table 2 and 3) for the performance of the proposed method. Our results showed greater DC value for prostate, bladder and rectum compared to Men *et al.*'s report. Gao *et al.* showed average DC values of 0.87, 0.92, and 0.88 and ASHD values of 1.77, 1.37 and 1.38 mm for prostate, bladder, and rectum, respectively[33]. Our model achieved equivalent or improved values for average DC and ASHD for all ROIs with the most improvement shown in rectum segmentation. More accurate segmentation of sensitive surrounding OARs such as the rectum is a critical component in the clinical application of auto-segmentation methods for minimizing radiation toxicity.

## 5. Conclusion

In summary, we showed that the U-Net structure can be applied successfully to male pelvic segmentation. The proposed method is highly accurate and reproducible for automated ROI segmentation, providing the basis to improve automatic delineation of the boundaries between target and surrounding normal soft tissues on a standard radiation therapy planning CT scan. This study could provide a foundation for improving accuracy and reproducibility of a critical step in radiation delivery to control disease while limiting adverse effects related to random errors in delineating ROIs in prostate cancer patients.

## Acknowledgements

AB and SJ would like to thank the Cancer Prevention and Research Institute of Texas (CPRIT) for their financial support through grant IIRA RP150485. The authors thank Dr. Damiana Chiavolini for editing the manuscript.

**References:**

1. van Mourik AM, Elkhuizen PH, Minkema D, Duppen JC, Dutch Young Boost Study G, van Vliet-Vroegindeweij C. Multiinstitutional study on target volume delineation variation in breast radiotherapy in the presence of guidelines. *Radiotherapy and oncology : journal of the European Society for Therapeutic Radiology and Oncology.* 2010;94(3):286-291.
2. Van Herk M. Errors and margins in radiotherapy. *Seminars in radiation oncology.* 2004;14(1):52-64.
3. Weiss E, Hess CF. The impact of gross tumor volume (GTV) and clinical target volume (CTV) definition on the total accuracy in radiotherapy. *Strahlentherapie und Onkologie.* 2003;179(1):21-30.
4. Weiss E, Richter S, Krauss T, et al. Conformal radiotherapy planning of cervix carcinoma: differences in the delineation of the clinical target volume: A comparison between gynaecologic and radiation oncologists. *Radiotherapy and Oncology.* 2003;67(1):87-95.
5. Rasch C, Steenbakkers R, van Herk M. Target Definition in Prostate, Head, and Neck. *Seminars in Radiation Oncology.* 2005;15(3):136-145.
6. Vorwerk H, Beckmann G, Bremer M, et al. The delineation of target volumes for radiotherapy of lung cancer patients. *Radiotherapy and Oncology.* 2009;91(3):455-460.
7. Geets X, Daisne JF, Arcangeli S, et al. Inter-observer variability in the delineation of pharyngo-laryngeal tumor, parotid glands and cervical spinal cord: Comparison between CT-scan and MRI. *Radiotherapy and Oncology.* 2005;77(1):25-31.
8. Jensen NKG, Mulder D, Lock M, et al. Dynamic contrast enhanced CT aiding gross tumor volume delineation of liver tumors: An interobserver variability study. *Radiotherapy and Oncology.* 2014;111(1):153-157.
9. Kepka L, Bujko K, Garmol D, et al. Delineation variation of lymph node stations for treatment planning in lung cancer radiotherapy. *Radiotherapy and Oncology.* 2007;85(3):450-455.
10. Steenbergen P, Haustermans K, Lerut E, et al. Prostate tumor delineation using multiparametric magnetic resonance imaging: Inter-observer variability and pathology validation. *Radiotherapy and Oncology.* 2015;115(2):186-190.
11. Steenbergen P, Haustermans K, Lerut E, et al. Prostate tumor delineation using multiparametric magnetic resonance imaging: Inter-observer variability and pathology validation. *Radiother Oncol.* 2015;115(2):186-190.
12. Njeh C. Tumor delineation: The weakest link in the search for accuracy in radiotherapy. *Journal of medical physics/Association of Medical Physicists of India.* 2008;33(4):136.
13. Zelefsky MJ, Fuks Z, Hunt M, et al. High-dose intensity modulated radiation therapy for prostate cancer: early toxicity and biochemical outcome in 772 patients. *Int J Radiat Oncol Biol Phys.* 2002;53(5):1111-1116.
14. Sander L, Langkilde NC, Holmberg M, Carl J. MRI target delineation may reduce long-term toxicity after prostate radiotherapy. *Acta Oncol.* 2014;53(6):809-814.
15. Greenham S, Dean J, Fu CK, et al. Evaluation of atlas-based auto-segmentation software in prostate cancer patients. *J Med Radiat Sci.* 2014;61(3):151-158.
16. Han X, Hoogeman MS, Levendag PC, et al. Atlas-Based Auto-segmentation of Head and Neck CT Images. 2008; Berlin, Heidelberg.

17. Isambert A, Dhermain F, Bidault F, et al. Evaluation of an atlas-based automatic segmentation software for the delineation of brain organs at risk in a radiation therapy clinical context. *Radiother Oncol.* 2008;87(1):93-99.
18. Teguh DN, Levendag PC, Voet PW, et al. Clinical validation of atlas-based auto-segmentation of multiple target volumes and normal tissue (swallowing/mastication) structures in the head and neck. *Int J Radiat Oncol Biol Phys.* 2011;81(4):950-957.
19. Korsager AS, Carl J, Riis Ostergaard L. Comparison of manual and automatic MR-CT registration for radiotherapy of prostate cancer. *J Appl Clin Med Phys.* 2016;17(3):294-303.
20. Rasch C, Barillot I, Remeijer P, Touw A, van Herk M, Lebesque JV. Definition of the prostate in CT and MRI: a multi-observer study. *Int J Radiat Oncol Biol Phys.* 1999;43(1):57-66.
21. Ramesh N YJ, Sethi IK. Thresholding based on histogram approximation. *IEEE Proc Vision Image Signal Proc.* 1995;142:271-279.
22. Sharma N, Aggarwal LM. Automated medical image segmentation techniques. *J Med Phys.* 2010;35(1):3-14.
23. Qian X, Wang J, Guo S, Li Q. An active contour model for medical image segmentation with application to brain CT image. *Med Phys.* 2013;40(2):021911.
24. Huang J, Jian F, Wu H, Li H. An improved level set method for vertebra CT image segmentation. *Biomed Eng Online.* 2013;12:48.
25. Osher SS, J. A. Fronts propagating with curvature-dependent speed: Algorithms based on Hamilton–Jacobi formulations. *J Comput Phys.* 1988;79:12-49.
26. Yezzi A, Jr., Kichenassamy S, Kumar A, Olver P, Tannenbaum A. A geometric snake model for segmentation of medical imagery. *IEEE Trans Med Imaging.* 1997;16(2):199-209.
27. Feulner J ZSK, Hammon M, Seifert S, Huber M, Comaniciu D, Hornegger J and Cavallaro A. A probabilistic model for automatic segmentation of the esophagus in 3D CT scans. *IEEE Trans Med Imaging.* 2011;30:1252-1264.
28. Yang J, Haas B, Fang R, et al. Atlas ranking and selection for automatic segmentation of the esophagus from CT scans. *Phys Med Biol.* 2017. doi: 10.1088/1361-6560/aa94ba.
29. Kirisli HA, Schaap M, Klein S, et al. Evaluation of a multi-atlas based method for segmentation of cardiac CTA data: a large-scale, multicenter, and multivendor study. *Med Phys.* 2010;37(12):6279-6291.
30. Sjoberg C, Lundmark M, Granberg C, Johansson S, Ahnesjo A, Montelius A. Clinical evaluation of multi-atlas based segmentation of lymph node regions in head and neck and prostate cancer patients. *Radiat Oncol.* 2013;8:229.
31. Wu M, Rosano C, Lopez-Garcia P, Carter CS, Aizenstein HJ. Optimum template selection for atlas-based segmentation. *Neuroimage.* 2007;34(4):1612-1618.
32. Aljabar P, Heckemann RA, Hammers A, Hajnal JV, Rueckert D. Multi-atlas based segmentation of brain images: atlas selection and its effect on accuracy. *Neuroimage.* 2009;46(3):726-738.
33. Gao Y, Shao Y, Lian J, Wang AZ, Chen RC, Shen D. Accurate Segmentation of CT Male Pelvic Organs via Regression-Based Deformable Models and Multi-Task Random Forests. *IEEE Trans Med Imaging.* 2016;35(6):1532-1543.
34. Lay N, Birkbeck N, Zhang J, Zhou SK. Rapid Multi-organ Segmentation Using Context Integration and Discriminative Models. 2013; Berlin, Heidelberg.

35. Shao Y, Gao Y, Wang Q, Yang X, Shen D. Locally-constrained boundary regression for segmentation of prostate and rectum in the planning CT images. *Med Image Anal.* 2015;26(1):345-356.
36. Shi Y, Gao Y, Liao S, Zhang D, Gao Y, Shen D. Semi-automatic segmentation of prostate in CT images via coupled feature representation and spatial-constrained transductive lasso. *IEEE Trans Pattern Anal Mach Intell.* 2015;37(11):2286-2303.
37. Krizhevsky A SI, Hinton GE. Imagenet classification with deep convolutional neural networks. *Adv Neural Inf Process Syst.* 2012;25:1106–1114.
38. Long J SE, Darrell T. Fully convolutional networks for semantic segmentation. *IEEE Trans Pattern Anal Mach Intell.* 2017;39:640-651.
39. Simonyan K ZA. Very deep convolutional networks for largescale image recognition. *Eprint Arxiv.* 2014.
40. Shiwen Shen AATB, JasonCong, WilliamHsu. An automated lung segmentation approach using bidirectional chain codes to improve nodule detection accuracy. *Computers in Biology and Medicine.* 2015;57:139-149.
41. Nanthagopal AP RR. A region-based segmentation of tumour from brain CT images using nonlinear support vector machine classifier. *J Med Eng Technol.* 2012;36:271-277.
42. Betancur J, Rubeaux M, Fuchs TA, et al. Automatic Valve Plane Localization in Myocardial Perfusion SPECT/CT by Machine Learning: Anatomic and Clinical Validation. *J Nucl Med.* 2017;58(6):961-967.
43. Ibragimov B, Xing L. Segmentation of organs-at-risks in head and neck CT images using convolutional neural networks. *Med Phys.* 2017;44(2):547-557.
44. Men K, Dai J, Li Y. Automatic segmentation of the clinical target volume and organs at risk in the planning CT for rectal cancer using deep dilated convolutional neural networks. *Med Phys.* 2017. doi: 10.1002/mp.12602.
45. Shelhamer E, Long J, Darrell T. Fully Convolutional Networks for Semantic Segmentation. *IEEE Trans Pattern Anal Mach Intell.* 2017;39(4):640-651.
46. Koltun FYaV. Multi-Scale Context Aggregation by Dilated Convolutions. 2015.
47. V. Badrinarayanan AKaRC. SegNet: A Deep Convolutional Encoder-Decoder Architecture for Scene Segmentation. *IEEE Transactions on Pattern Analysis & Machine Intelligence.* 2017;99.
48. Ronneberger O, Fischer, P., Brox, T. U-Net: Convolutional Networks for Biomedical Image Segmentation. *Medical Image Computing and Computer-Assisted Intervention-MICCAI.* 2015. doi: 10.1007/978-3-319-24574-4_28:234-241.
49. Sergey Ioffe CS. Batch Normalization: Accelerating Deep Network Training by Reducing Internal Covariate Shift. *arXiv.* 2015;1502.03167.
50. G. E. Hinton NS, A. Krizhevsky, I. Sutskever and R. R. Salakhutdinov. Improving neural networks by preventing co-adaptation of feature detectors. *arXiv.* 2012;207.058.
51. Chollet Fccoao. Keras. 2015.
52. Milletari F, Navab N, Ahmadi S-A. V-Net: Fully Convolutional Neural Networks for Volumetric Medical Image Segmentation. *ArXiv e-prints.* 2016;1606. http://adsabs.harvard.edu/abs/2016arXiv160604797M Accessed June 1, 2016.
53. Kaiming He XZ, Shaoqing Ren, Jian Sun. Deep Residual Learning for Image Recognition. *ArXiv.* 2015.

54. Oscar Acosta JD, Gael Drean, Antoine Simon, Renaud de Crevoisier, Pascal Haigron. Multi-Atlas-Based Segmentation of Pelvic Structures from CT Scans for Planning in Prostate Cancer Radiotherapy. *Abdomen and Thoracic Imaging.* 2013.623-656.
55. Warfield SK, Zou KH, Wells WM. Simultaneous truth and performance level estimation (STAPLE): an algorithm for the validation of image segmentation. *IEEE Trans Med Imaging.* 2004;23(7):903-921.
56. Rohlfing T, Russakoff DB, Maurer CR. Expectation maximization strategies for multi-atlas multi-label segmentation. *Inf Process Med Imaging.* 2003;18:210-221.